\documentstyle[12pt]{article}
\input psfig.sty

\newcommand{\beq}{\begin{equation}}
\newcommand{\eeq}[1]{\label{#1} \end{equation}}
\newcommand{\insertplot}[1]{\centerline{\psfig{figure={#1},width=13.3cm}}}
\newcommand{\insertplotshort}[1]{\centerline{\psfig{figure={#1},height=7.0cm}}}
\newcommand{\insertplotsshort}[1]{\centerline{\psfig{figure={#1},height=5.0cm}}}
\newcommand{\insertplotl}[1]{\centerline{\psfig{figure={#1},height=9.0cm}}}
\newcommand{\insertplotll}[1]{\centerline{\psfig{figure={#1},height=11.0cm}}}
\newcommand{\insertplotlll}[1]{\centerline{\psfig{figure={#1},height=19.0cm}}}

\title{\bf Duality
 in strong interactions\footnote{Lecture delivered by L. Jenkovszky at the 6th 
 International Summer School-Seminar On Actual Problems
 Of High Energy Physics, Gomel, Belarus, August 7-16, 2001.}
 }
\author{L. Jenkovszky$^{1\star}$, V.K. Magas$^{1,2\dagger}$ and E.
Predazzi$^{3\diamond}$\\  \\
{\it $^1$~Bogolyubov Institute for Theoretical Physics}\\
{\it Academy of Sciences of Ukraine}\\
{\it 01143 Kiev, Ukraine} \\  
{\it $^2$~Center for Physics of Fundamental Interactions}\\ 
{\it Instituto Superior Tecnico}\\
{\it Av. Rovisco Pais, 1049-001 Lisbon, Portugal}\\
{\it $^3$Torino University and INFN}\\  
{\it via P. Giuria, 1, Torino, 10125 Italy}\\
{\it $^{\star}$~Email: jenk@gluk.org}\\
{\it $^{\dagger}$~Email: vladimir@cfif.ist.utl.pt}\\
{\it $^{\diamond}$~Email: predazzi@to.infn.it}
}

\begin{document}

\maketitle


    In this lecture we first present the classical results on duality in
the strong interactions, developed in the late 60-es and the early 70-es, 
regarding on-mass-shell hadronic reactions in the framework of the analytical 
$S-$matrix theory.  Then we discuss  their extension to of-mass-shell processes 
(virtual Compton- and deep inelastic scattering), published in a series of 
recent papers \cite{FJM,JM,JMP01}  
 
    The continuity of dynamics in going from low to high energies, from low
to high virtualities, from exclusive to inclusive processes etc is a basic 
principle of the strong interactions (see \cite{BK}). Duality between 
Reggeon-resonance is a particularly nice and instructive example showing how it 
works. Before its  discovery, the contribution from low-energy, direct channel 
resonances was considered to be independent from that of high-energy Regge 
exchanges, and the complete scattering amplitude was thought to be their sum. 
This was the interference model. Later on, in analyzing finite energy sum rules, 
it was realized that each piece already contains the whole dynamics and 
that a proper sum of direct channel resonances "knows" its high energy, Regge  
behavior and vice verse: from a smooth Regge asymptotic formula resonances 
can be created and thus the sum of $s$-channel resonances and $t$-channel 
exchanges (interference model) leads to wrong "double counting". Instead, 
their equivalence, or duality proved to be the right answer. In a similar way,
below we shall ask in which way the small- and large-$x$ parts of the structure 
functions are related.

An  important further step in the late 60-es consisted in the construction of 
explicit models realizing duality. The Veneziano model \cite{Veneziano} is the 
best known example. The common feature of most of the 
dual models was their "soft" dynamics, which means that they were applicable 
only for peripheral ("soft") collisions and did not contain any information on
the interaction at short distances, where the partonic nature of the strong 
interaction may reveal. The reason for this shortcoming was in the linear 
character of the Regge trajectories admitted in those "narrow resonance" dual 
models. Their contradiction with analiticity and unitarity and the absence of
any imaginary part was the main reason why the physical applications of
dual models were abandoned, giving place to formal mathematical developments. 

A revival of this subject was partly due to the recent experimental measurements 
of the nucleon structure function at JLab (CEBAF) \cite{Niculescu}.  They 
inspired the present authors in the construction of a unified 
"two-dimensionally dual" picture of the strong
interaction \cite{FJM,JM,JMP01} connecting low- and high energies (Veneziano, or 
resonance-reggeon duality \cite{Veneziano}) with low- and high 
virtualities ($Q^2$)
(Bloom-Gilman, or hadron-parton duality \cite{BG}). 
The basic idea of the unification is the use of a
$Q^2$-dependent dual amplitudes,
employing nonlinear complex Regge trajectories providing an
imaginary part in the scattering amplitude, related to the total
cross section and structure functions and thus saturating duality by a
finite number of resonances lying on the (limited) real part of
the Regge trajectories.

\begin{figure}[htb]
        \insertplot{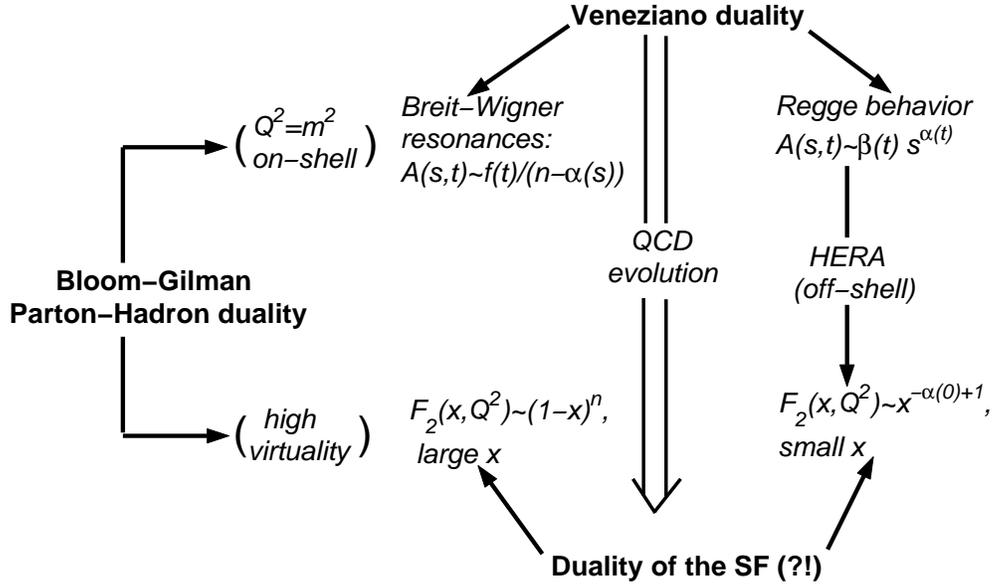}
\caption{Veneziano, or resonance-reggeon
duality \cite{Veneziano} and
Bloom-Gilman, or hadron-parton duality \cite{BG}
 in strong interactions} 
\label{diag}
\end{figure}

 The resulting object, a deeply virtual
scattering amplitude, $A(s,t,Q^2)$, is a function of three
variables, reducing to a nuclear structure function (SF) when
$t=0$ and to an on-shell hadronic scattering amplitude for
$Q^2=m^2$.  It closes the circle in Fig \ref{diag}.
 We use this amplitude to describe the 
background as well as the resonance component \cite{FH}.

The $Q^2-$ dependence of the residuae functions here will be
chosen in such a way as to provide for Bjorken scaling at small
$x$ (large $s$). The resulting amplitude (structure function) is
applicable in the whole kinematical range, including the resonance
region. We call this unification "two dimensional duality" - one
in $s$, the other one in $Q^2$,

In the early days of duality, off mass continuation was attempted
\cite{RR} by means of multi-leg (e.g. 6-point) dual amplitudes
with "extra" lines taken at their poles. Without going into
details, here we only mention that scaling in this approach can be
achieved \cite{Schierholz} only with nonlinear trajectories, e.g.
those with logarithmic or constant asymptotic.

\section {Notation and conventions}
\begin{figure}[htb]
        \insertplotsshort{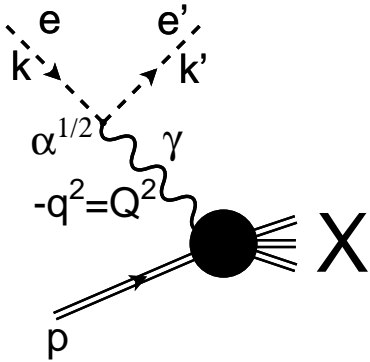}
\caption{Kinematic of deep inelastic scattering.} 
\label{d1}
\end{figure}

We use standard notation for the cross section and structure
function (see Fig. \ref{d1}):
 \beq
 \sigma^{\gamma^* p}={4\pi^2\alpha(1+4m^2x^2/Q^2
)\over{Q^2(1-x)}}{F_2(x,Q^2)\over{1+R(x,Q^2)}}, \eeq{eq1} where
$\alpha$ is the fine structure constant, $Q^2$ is minus the
squared four momentum transfer or the momentum carried by the
virtual photon, $x$ is the Bjorken variable and $s$ is the squared
center of mass energy of the $\gamma^*p$ system, obeying the
relation \beq s=Q^2(1-x)/x+m_p^2, \eeq{eq2} where $m_p$ is the
proton mass and $R(x,Q^2)= \sigma_L(x,Q^2)/\sigma_T(x,Q^2)$. For
the sake of simplicity we set $R=0,$ which is a reasonable
approximation.

\begin{figure}[htb]
        \insertplot{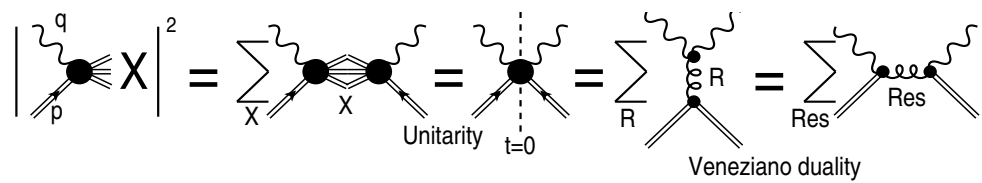}
\caption{According to the Veneziano (or resonance-reggeon) duality a proper sum 
of either t-channel or s-channel resonance exchanges accounts for
the whole amplitude. 
{\it Diagram drawings in Figs. \ref{d1} and  \ref{d2} have been 
made with {\bf scribble1.2} program \cite{scribb}.}
} 
\label{d2}
\end{figure}

We use the norm where \beq \sigma_T^{\gamma^*}(s,t,Q^2)=Im\
A(s,t,Q^2). \eeq{eq3} According to the two-component duality
picture \cite{FH} both the scattering amplitude $A$ and the structure
function $F_2$ are sums of a diffractive and non-diffractive
terms. At high energies both terms are Regge-behaved. In $\gamma^*
p$ scattering only positive signature exchange are allowed. The
dominant ones are the Pomeron and the $f$ Reggeon, respectively.
The relevant scattering amplitude is (remember that here $t=0$)
\beq
A_i(s,Q^2)=i\beta_k(Q^2)\Bigl(-i{s\over{s_i}}\Bigr)^{\alpha_k(0)-1},
\eeq{eq4} where $\alpha$ and $\beta$ are the Regge trajectory and
residue and $k$ stands either for the Pomeron or the Reggeon. As usual, the
residue will be chosen such as to satisfy approximate Bjorken
scaling for the structure function \cite{BGP,K}. It should be
remembered that by factorization, assuming that the Reggeon (or
Pomeron) exchange is a simple pole, the residue function is a
product of two vertices - the $\gamma\gamma R(P)$ and $NNR(P)$,
where $N$ stands for the nucleon (see Fig. \ref{d2}).

At low energies the scattering amplitude is dominated by the
contribution of the near-by resonances. In the vicinity of a resonance
$Res$, the amplitude can be also written in a factorized form,
product of the probabilities that two particles - $\gamma$ and $p$
form a resonance with squared mass $s_R$ and total width $\Gamma$
\beq
A(s,Q^2)=\sum_{spin}{A_{fi}(Q^2)A^*_{if}(Q^2)\over{s_{Res}-s-i\Gamma}},
\eeq{eq5} where the $s_R$ is the squared mass of the resonance and
$\Gamma(Q^2)$ is its width; $A_{fi}$ are the inelastic form
factors.

\section{Nucleon resonances in inelastic electron-nucleon
scattering}
Some thirty years ago Bloom and Gilman \cite{BG}
observed that the prominent resonances in  inelastic
electron-proton scattering do not disappear with increasing $Q^2$
relatively to the "background" but instead fall at roughly the
same rate as any background. Furthermore, the smooth scaling limit
proved to be an accurate average over resonance bumps seen at
lower $Q^2$ and $s$.

\begin{figure}[htb]
        \insertplotll{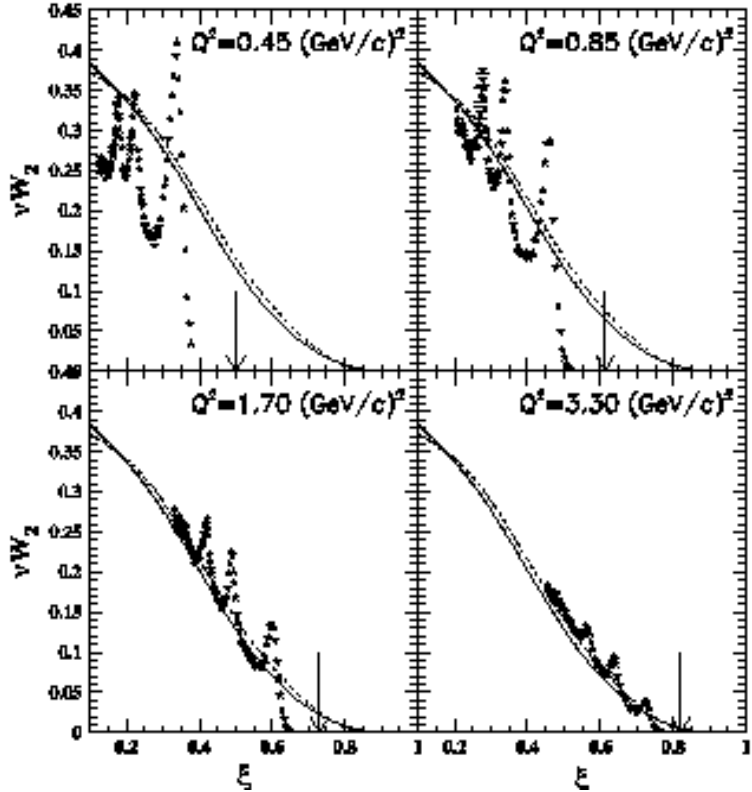}
\caption{Extracted $\nu W_2$ structure function 
spectra as a function of 
Nachtmann scaling variable (\ref{nach}).
The solid curve indicates a fit to DIS data for $Q^2=10\ GeV^2$,
dashed curve - $Q^2=5\ GeV^2$. From \cite{Niculescu}.} 
\label{nicul}
\end{figure}

Since then, the phenomenon was studied in a number of papers
\cite{carlson,stoler,Carl} and recently has been confirmed experimentally
\cite{Niculescu} (see Fig. \ref{nicul}). These studies were aimed 
mainly to answer the
questions: in which way a  limited number of resonances can reproduce 
the smooth scaling behavior? The main
theoretical tools in these studies were finite energy sum rule
and perturbative QCD calculations, whenever applicable. Our aim
instead is the construction of an explicit dual model combining
direct channel resonances, Regge behavior, typical of hadrons and
scaling behavior, typical of the partonic picture. 

The  existence of resonances in the structure function at large
$x$, close to $x=1$ by itself is not surprising: by the relations
(\ref{eq1}) and (\ref{eq2}) they are the same as in $\gamma^*p$
total cross section, seen in different variables. The important
question is if and how a small number  of resonances (or even a
single one) can reproduce the smooth Bjorken scaling behavior,
known to be an asymptotic property, typical of multiparticle
processes.

The possibility that a limited (small) number of resonances can
build up smooth Regge behavior was demonstrated by means of finite
energy sum rules \cite{DHS}. Further it was confused by the presence of an
infinite number narrow resonances in the Veneziano model
\cite{Veneziano}, which made its phenomenological application
difficult, if not impossible. Similar to the case of the
resonance-reggeon duality \cite{DHS}, hadron-parton duality was
established \cite{BG} by means of the finite energy sum rules, 
but it was not realized explicitly like the Veneziano
model (or its further modifications).

Actually, the early onset of Bjorken scaling, called "early, or
precaution scaling" was observed with the first measurements of
deep inelastic scattering at SLAC, where it was noticed that a
more rapid approach to scaling can be achieved with the BG
variable \cite{BG} $x'=x/(1+m^2x^2/Q^2)$ instead of $x$ (or
$\omega=1/x$). More recently the following generalization of the
BG variable
\beq
\xi={2x\over{1+\sqrt{1+{4m^2x^2\over Q^2}}}}
\eeq{nach}
was suggested by O.Nachtmann \cite{Nachtmann}. We use the standard 
Bjorken variable $x$, however our results can be easily rewritten in terms
of the above-mentioned modified variables.

First attempts to combine resonance (Regge) behavior with
Bjorken scaling were made \cite{DG,BEG,EM} at low energies (large
$x$), with the emphasis on the right choice of the
$Q^2$-dependence, such as to satisfy the required behavior of the
form factors, vector meson dominance (VMD) with the requirement of the 
Bjorken scaling. (N.B.: the
validity (or failure) of the (generalized) VMD
is still disputed). Similar attempts in the high-energy (low $x$)
region became popular recently, with the advent of the HERA data.
They will be presented in Sec. \ref{q2dep}.

A consistent treatment of the problem requires the account for the
spin dependence. For simplicity we ignore it in this lecture (see e.g.
\cite{Carl}).

\section{Factorization and dual properties (bootstrap)
of the vertices}
Let us remind that the residue functions are completely arbitrary
in the Regge pole model, but they are constrained in the dual
model. We show this by using the low energy- (resonances) and high
energy- (Regge) decomposition on the simple Veneziano model
\cite{Veneziano} $$V(s,t)=\int_0^1 dz
z^{-\alpha(s)}(1-z)^{-\alpha(t)}= $$ \beq
B(1-\alpha(s),1-\alpha(t)) =
{\Gamma(1-\alpha(s))\Gamma(1-\alpha(t))
\over{\Gamma(2-\alpha(s)-\alpha(t))}}. \eeq{eq11} \beq
V(s,t)=\sum_{n=1}^{\infty}{1\over{n-\alpha(s)}}{\Gamma(n+\alpha(t)+1)
\over {n!\ \Gamma(\alpha(t)+1)}}. \eeq{eq12}

By the Stirling formula
$$
 V(s,t)\Bigg|_{|\alpha(s)|\rightarrow\infty}\rightarrow
[-\alpha(s)]^ {\alpha(t)-1} \Gamma\bigl(1-\alpha(t)\bigr)
$$
\beq
 \biggl[\sum_{n=0}^N{a_n(0) \over
{[\alpha(s)]^n}}+0\biggl({1\over{[\alpha(s)]^{N+1}}}\biggr)
\biggr] \eeq{eq13} and since for small $|t|$ the $\Gamma$ function
varies slowly compared with the exponential, the Regge asymptotic
behavior is \beq V(s,t)\sim (\alpha' s)^{\alpha(t)}, \eeq{eq14}
where $\beta(t)=(\alpha')^{\alpha(t)}$ is the Regge residue.

Actually, one has to identify a single (and hence economic!) Regge
exchange amplitude with a sum of direct channel poles. Such an
identification is not practical for an infinite number of poles
(e.g. the Veneziano amplitude) but, as we show below is feasible
if their number is finite (small). To anticipate the forthcoming
discussion, we shall feed the $Q^2$-dependence in the Regge
residue at high energies (small $x$ and use the dual amplitude
with a finite number of resonances!) to the whole kinematical
region, including that of resonances. Relating the amplitude to
the SF, we set $t=0.$

To remedy the problems of the infinite number of narrow resonance,
nonunitarity and lack of an imaginary part, we use a
generalization of the Venezinano model free from the
above-mentioned difficulties.

\section{Dual amplitude with Mandelstam analyticity }
The invariant dual on-shell scattering
amplitude dual amplitude with Mandelstam analyticity (DAMA)
applicable  both to the diffractive and non-diffractive components
reads \cite{D} \beq D(s,t)=\int_0^1 {dz \biggl({z \over g}
\biggr)^{-\alpha(s')-1} \biggl({1-z \over
g}\biggr)^{-\alpha(t')-1}}, \eeq{eq21} where $s'=s(1-z), \ \
t'=tz, \ \ g$ is a parameter, $g>1$, and $s, \ \ t$ are the
Mandelstam variables.

For $s\rightarrow\infty$ and fixed $t$ it has the following Regge
asymptotic behavior \beq
D(s,t)\approx\sqrt{{2\pi\over{\alpha_t(0)}}}g^{1+a+ib}\Biggl({s\alpha'(0)g\ln
g\over{\alpha_t(0)}}\Biggl)^{\alpha_t(0)-1}, \eeq{eq22} where
$a=Re\ \alpha\Bigl({\alpha_t(0)\over{\alpha'(0)\ln g}}\Bigr)$ and
$b=Im\ \alpha\Bigl({\alpha_t(0) \over{\alpha'(0)\ln g}}\Bigr)$.

The pole structure of DAMA is similar to that of the Veneziano
model except that multiple poles may appear at daughter levels.
The presence of these multipoles does not contradict the
theoretical postulates.  On the other hand, they can be removed
without any harm to the dual model by means the so-called Van der
Corput neutralizer. The procedure  \cite{D} is to multiply the
integrand of (\ref{eq21}) by a function $\phi(x)$ with the
properties:
$$ \phi(0)=0,\ \ \ \phi(1)=1,\ \ \ \phi^n(1)=0,\ \ n=1,2,3,... $$
The function $ \phi(x)=1-exp\Biggl({-{x\over{1-x}}}\biggr), $ for
example, satisfies the above conditions and results \cite{D} in a
standard, "Veneziano-like" pole structure: \beq
D(s,t)=\sum_ng^{n+\alpha_t(0)}{C_n\over{n-\alpha(s)}}, \eeq{eq23}
where \beq
C_n={\alpha_t(0)\Bigl(\alpha_t(0)+1\Bigr)...\Bigl(\alpha_t(0)+n+1\Bigr)\over{n!}}.
\eeq{eq24}

The pole term (\ref{eq23}) is a generalization of the Breit-Wigner
formula (\ref{eq5}), comprising a whole sequence of resonances
lying on a complex trajectory $\alpha(s)$. Such a "reggeized"
Breit-Wigner formula has little practical use in the case of
linear trajectories, resulting in an infinite sequence of poles,
but it becomes a powerful tool if complex trajectories with a
limited real part and hence a restricted number of resonances are
used. Moreover, it appears that a small number of resonances is sufficient
to saturate the direct channel.

Contrary to the Veneziano model, DAMA (\ref{eq21}) not only
allows but rather requires the use of nonlinear complex
trajectories providing, in particular, for the imaginary part of
the amplitude, resonance widths and resulting in a finite number
of those. More specifically, the asymptotic rise of the
trajectories in DAMA is limited by the condition (in accordance
with an important upper bound derived earlier \cite{DP}) \beq
|{\alpha(s)\over{\sqrt s\ln s}}|\leq const, \ \
s\rightarrow\infty. \eeq{lim} Actually, this upper bound can be
even lowered up to a logarithm by requiring wide angle scaling
behaviour for the amplitude.

Models of Regge trajectories combining the correct threshold and
asymptotic behaviors have been widely discussed in the literature
(see e.g. \cite{FJMPP} for a recent treatment of this problem). A
particularly simple model is based on a sum of square roots 
$$\alpha(s)=\alpha_0+\sum_i\gamma_i \sqrt{s_i-s}, $$
 where the
lightest threshold (made of two pions or a pion and a nucleon) is
important for the imaginary part, while the heaviest threshold
limits the rise of the real part, where resonances terminate.

DAMA with the trajectories specified above is equally applicable
to both: the diffractive and non-diffractive components of the
amplitude, the difference being qualitative rather than
quantitative. The utilization of a trajectory with a single
threshold, \beq
\alpha_E(s)=\alpha_E(0)+\alpha_{1E}(\sqrt{s_E}-\sqrt{s_E-s})
\eeq{eq32} prevents the production of resonances on the the
physical sheet, although they are present on the nonphysical
sheet, sustaining duality (i.e. their sum produces Regge
asymptotic behavior). This nontrivial property of DAMA makes it
particularly attractive in describing the smooth background (dual
to the Pomeron exchange) (see \cite{D}). The threshold value, slope and the
intercept of this exotic trajectory are free parameters.

For the resonance component a finite sum  in (\ref{eq23}) is
adequate, but we shall use a simple model with lowest
threshold included explicitly and the higher ones approximated by
a linear term: \beq
\alpha_R(s)=\alpha_R(0)+\alpha's+\alpha_{1R}(\sqrt{s_0}-\sqrt{s_0-s}),
\eeq{eq33} where $s_0$ is the lowest threshold --
$s_0=(m_{\pi}+m_p)^2$ in our case -- while the remaining 3
parameters will be adjusted to the known properties of the
relevant trajectories (those of $N^*$ and the $\Delta$ isobar in our case).
 The termination of resonances, provided in DAMA by the limited
real part, here will be effectively taken into account by a cutoff
in the summation of (\ref{eq23}).

Finally, we note that a minimal model for the scattering amplitude
is a sum \beq A(s,t,u)=c(D(s,t)+D(u,t)), \eeq{eq34} providing 
the correct signature at high-energy limit, $c$ is a normalization
factor. We disregard the symmetry (spin and isospin) properties of
the problem, concentrating on its dynamics. In the limit
$s\rightarrow\infty,\ \ t=0$ we have $u=-s$ and therefore \beq
A(s,0,-s)|_{s\rightarrow\infty}=c\ D(s,0)(1+(-1)^{\alpha_t(0)-1}),
\eeq{eq35a} where $D(s,t)$ is given by eq. (\ref{eq22}). For the
total cross section in this limit we obtain:
$$
\sigma_T^{\gamma^*}=Im\ A=Cg^{\alpha_t(0)+a} \left(s \alpha'(0)
\ln g \right)^{\alpha_t(0)-1}\cdot$$ \beq \cdot\left(\sin
(\alpha_t(0)-1)\pi \cos (b\ln g)+
(1+\cos (\alpha_t(0)-1)\pi)\sin (b\ln g) \right)\ ,\eeq{eq35}
where $C$ is a constant independent on $s,\ g$ and $\alpha'(0)$.

\section{$Q^2-$ dependence}
\label{q2dep}
Our main idea is the introduction of the  $Q^2$-dependence in the
dual model by matching its Regge asymptotic behavior and pole
structure to standard forms, known from the literature. The point
is that the correct identification of this $Q^2$-dependence in a
single asymptotic limit of the dual amplitude will extend it to
the rest of the kinematical regions. We have two ways to do so:\\
A) Combine  Regge behavior and Bjorken scaling limits of the
structure functions (or $Q^2$-dependent $\gamma^*p$ cross
sections)\\
B) Introduce properly $Q^2$ dependence in the resonance region.\\
They should match if the procedure is correct and the dual
amplitude should take care of any further inter- or
extrapolation.

\begin{figure}[htb]
        \insertplotshort{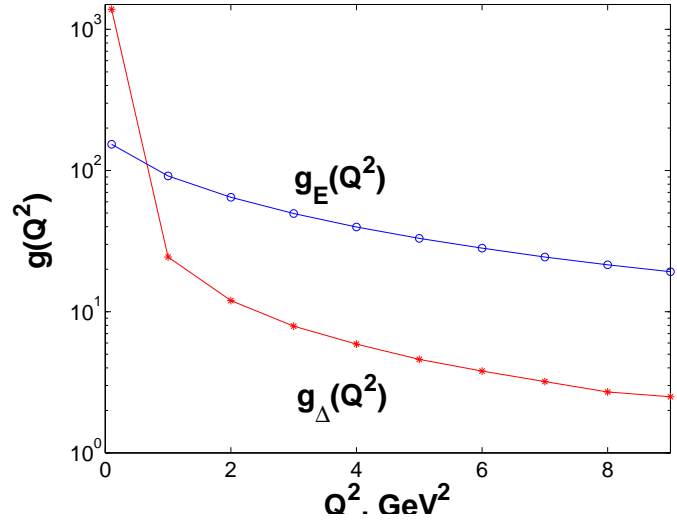}
\caption{$g(Q^2)$ - the solution of the transcendent equation 
(\ref{eq45}) - for $\Delta$ and exotic
trajectories.} 
\label{fig1}
\end{figure}

It is obvious from eq. (\ref{eq4}) that asymptotic Regge and
scaling behavior require the residue to fall like
$\sim(Q^2)^{-\alpha_i(0)+1}$. Actually, it could be more involved
if we require the correct $Q^2\rightarrow 0$ limit to be respected
and the observed scaling violation (the "HERA effect") to be
included. Various models to cope with the above requirements have
been suggested \cite{BGP,K,JMP99}. At HERA, especially at large
$Q^2$, scaling is so badly violated that it may not be explicit
anymore.

In combining Regge asymptotic behavior with (approximate) Bjorken
scaling, one can proceed basically in the following ways -- keep
explicitly a scaling factor $x^{\Delta}$ (to be broken by some
$Q^2$-dependence "properly" taken into account) \cite{K}

\beq F_2(x,Q^2)\sim
x^{-\Delta(Q^2)}\Bigl({Q^2\over{Q^2+Q_0^2}}\Bigr)^{1+\Delta(Q^2)},
\eeq{eq41} where $\Delta(Q^2)=\alpha_t(0)-1$ may be a constant, in
particular.

In the Regge asymptotic limit of the Veneziano model - 
$\sim(-\alpha' s)^{\alpha(t)-1},$ - $Q^2$-dependence can be 
introduced only  through the slope of the direct channel trajectory \cite{FJM}, i.e.
by making $\alpha'_s$ $Q^2$- dependent. This is a crucial point to be 
understood.  By Regge-factorization, the trajectory should not depend on the 
parameters (masses, virtualities) of the external particles. On the other hand, 
there is a nice bootstrap condition in dual models, by which one of the states 
on the trajectory (e.g. the lightest one) should be identified with 
the external particle, which means that the slope is associated with its squared 
mass. This can be generalized to a varying mass or $Q^2$. 

A different interpretation may come from treating a Regge pole as and effective 
one, accounting for unitarization etc and consequently embodying 
a complicated  $Q^2$-dependence. The latter interpretation became
popular \cite{BGP,K} in the applications of Regge formulae to explain the 
drastic increase of 
the structure functions with $Q^2$, observed at HERA 
(Bjorken scaling violation).
Although we do not
exclude this possibility (treating it as ``effective'' Regge pole, 
we study here the different option of introducing  scaling violation 
in the constant $g$ appearing, besides $\alpha'$, in the the residue 
of DAMA, eq (11).

From the explicit Regge asymptotic form of DAMA, (\ref{eq35}), and
neglecting the logarithmic dependence of $g$  we make the
following identification \beq
g(Q^2)^{\alpha_t(0)+a}=\left(Q_{lim}^2
\over{Q^2+Q_0^2}\right)^{\alpha_t(0)}. \eeq{eq45} 
Note that eq. (\ref{eq45}) is transcendent with
respect to $g$, since $a=a(g)= Re\
\alpha\Bigl({\alpha_t(0)\over{\alpha'(0)\ln g}}\Bigr)$. Another
point to mention is that this equation is not valid in the whole 
range of $Q^2$,
since for $Q^2$ close to $Q_{lim}^2$, $g$ may get smaller than $1$,
which is unacceptable in DAMA. For large $Q^2$ the
$Q^2$-dependence of the $\log g$ and $b=b(Q^2)=Im\
\alpha\Bigl({\alpha_t(0) \over{\alpha'(0)\ln g}}\Bigr)$ in eq.
(\ref{eq35}) can not be neglected, it might contribute to scaling
violation.

\begin{figure}[htb]
        \insertplotl{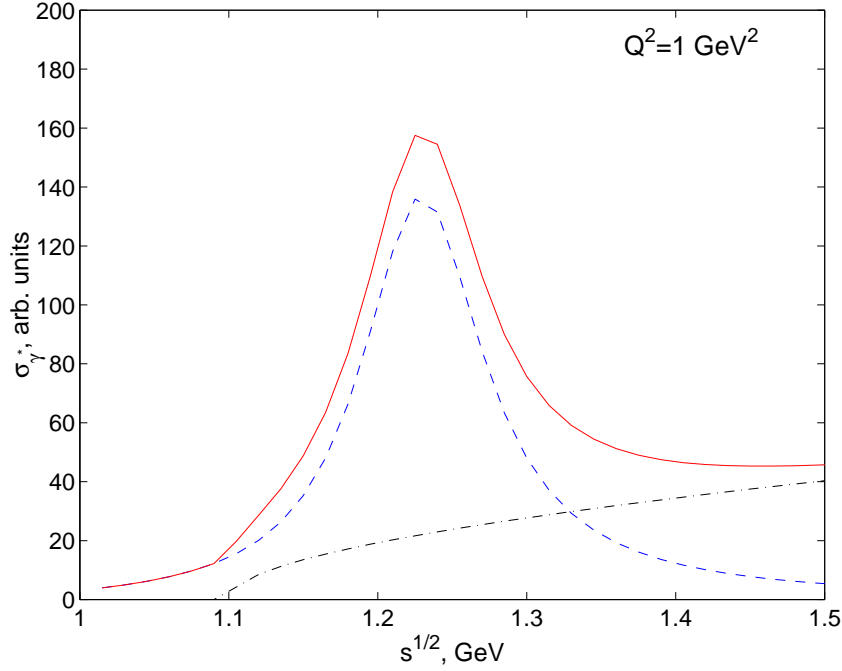}
\caption{ $\gamma^* p$ total cross section as a
function of $\sqrt{s}$. The dashed line shows the contribution
from the $\Delta$ resonance, the dot-dashed line corresponds to
the background, i.e. the contribution from the exotic trajectory.
Here $Q^2=1\ GeV^2$. } 
\label{fig2}
\end{figure}

\section{Scaling at large $x$}

Let us now consider the extreme case of a single resonance
contribution.

A resonance pole in DAMA contributes with 
$$A(s,t)=g^{n+\alpha_t(0)} {C_n\over{n-\alpha(s)}}.$$
At the resonance $s=s_{Res}$ one has $Re\ \alpha(s_{Res})=n$ and
${Q^2(1-x)\over x}=s_{Res}-m^2$, hence
$$
F_2(x,Q^2)={Q^2(1-x)\over{4\pi^2\alpha\Bigl(1+{4m^2x^2\over
{Q^2}}\Bigr)}} {C_n\over {Im\ \alpha(s_{Res})}}g(Q^2)^{n+\alpha_t(0)}.
$$
As $x\rightarrow 1$
$Q^2\approx{s_{Res}-m^2\over{1-x}}\rightarrow
\infty$ and
$$F_2(x,Q^2)\sim g\Bigl({s_{Res}-m^2\over{1-x}}\Bigr)^{n+\alpha_t(0)}.$$ 
By using the approximate solution
$
g(Q^2)\approx\left({Q^2_{lim}/
Q^2}\right)^{\alpha_t(0)\over{\alpha_t(0)+a}},
$
where $a$ is a slowly varying function of $g$, we get for $x$ near $1$
$$F_2(x,Q^2)\sim (1-x)^{\alpha_t(0)(n+\alpha_t(0))\over{\alpha_t(0)+a}},
$$
where the limits for $x$ are defined by
$Q_0^2\ll{s_{Res}-m^2\over{1-x}}\leq Q^2_{lim}.$

We recognize a typical large-$x$ scaling behavior $(1-x)^N$ with
the power $N$ (counting the quarks in the reaction)  depending
basically on the intercept of the $t$-channel trajectories.

\begin{figure}[htb]
        \insertplotl{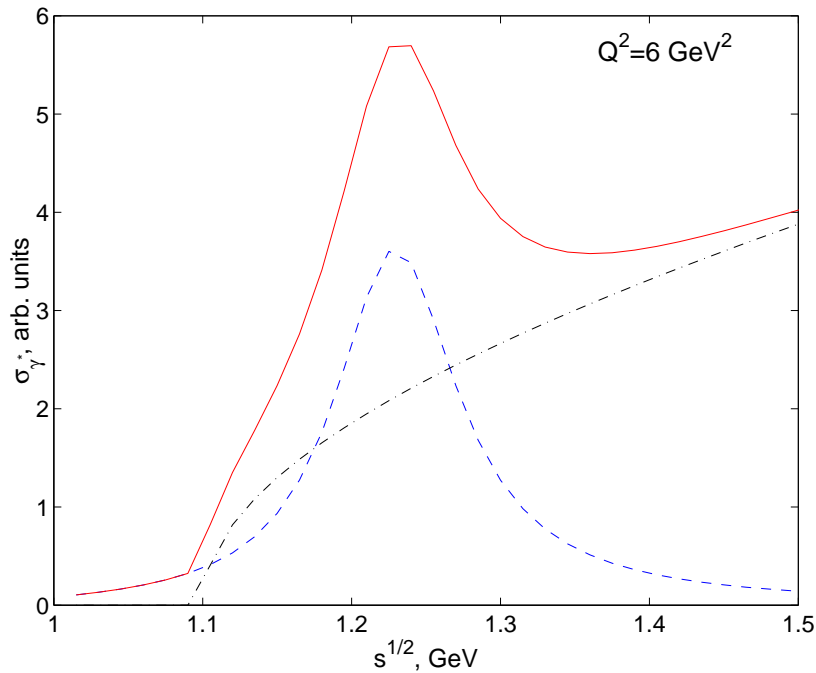}
\caption{The same as Fig. \ref{fig2}, but $Q^2=6\ GeV^2$. } 
\label{fig3}
\end{figure}

\section{Numerical Estimates}

Having fixed the $Q^2$-dependence of the dual model  by matching
its Regge asymptotic behavior with that of the structure
functions, we now use this dual model to extrapolate down to the
resonance region, where its pole expansion (\ref{eq23}) is
appropriate - now complemented with a $Q^2$ -dependence through
$g(Q^2)$, fixed by eq. (\ref{eq45})

\begin{table}[htb]
\caption{Values of parameters used in the calculations shown in 
Figs.  \ref{fig1},~\ref{fig2} and \ref{fig3}.} \label{tab1}
\medskip

\renewcommand{\arraystretch}{1.2} 
\begin{tabular}{|lll|}
\hline
\hfill \vline&$\Delta$ Resonance \hfill \vline & Background \\
\hline
$Q_{lim}^2,\ GeV^2$\hfill \vline& 62\hfill \vline  & 120\\
$Q_0^2,\ GeV^2$\hfill \vline&0.01\hfill \vline  & 2.5  \\
\hline
Dual \hfill \vline&$\alpha_f(t)$ is\hfill \vline &$\alpha_P(t)$ is\\
trajectory \hfill \vline&dual to $\alpha_{\Delta}(s)$\hfill \vline
&dual to $\alpha_E(s)$\\
\hline
\hfill \vline &$\alpha_f(0)=0.9$\hfill \vline & $\alpha_P(0)=1+0.077\cdot$\\
\hfill \vline &\hfill \vline  &$\cdot\left(1+\frac{2Q^2}
{Q^2+1.117}\right)$ \cite{K}\\
\hline
\end{tabular}\\

Normalization coefficient $c=0.03$.
\end{table}

As already said, we write the imaginary part of the scattering
amplitude as the sum of two terms - a diffractive (background) and
non-diffractive (resonance) one. Note that $g(Q^2)$ has the same
functional form (\ref{eq45}) in both cases, only the values of the
parameters differ (they are fixed from the small-$x$ fits
\cite{JMP99} of the SF).

At low, resonance, energies $\gamma^* p$ scattering exhibits a rich
resonance structure, intensively studied in a number of papers.
About 20 resonances overlap, their relative importance varying
with $Q^2$, but only a few can be identified more or less
unambiguously. These are: $\Delta^+(1236)$ with $J^P={3^+\over 2}$,
$N^{*+}(1520)$, $J^P={3^-\over 2}$, $N^{*+}(1688)$,
$J^P={5^+\over 2}$ and $N^{*+}(1920)$ with $J^P={7^+\over 2}$ (see Fig. \ref{ress}). They
lie on the $\Delta$ and the exchange-degenerate $N$ trajectories.
In this work we are mainly interested in introducing
$Q^2$-dependence into the scattering amplitude, therefore we will
concentrate on a single resonance ($\Delta^+(1236)$) at different
values of $Q^2$. We use trajectories (\ref{eq33}) in which the
lowest pion-nucleon threshold is included explicitly, while higher
thresholds are approximated by a linear term:
$$\alpha_{\Delta}(s)=0.1+0.84s+0.1331(\sqrt{s_0}-\sqrt{s_0-s}),$$
where $s_0=(m_\pi^2+m_N^2)$.\footnote{ Actually, trajectories
without any linear term (see e.g. \cite{FJMPP}) could be more
appropriate (and will be studied in future).} The above values of
the parameters are chosen so as to fit the known mass and width of the
$\Delta$ resonance in a way consistent with the known linear
parameterizations.

\begin{figure}[htb]
        \insertplot{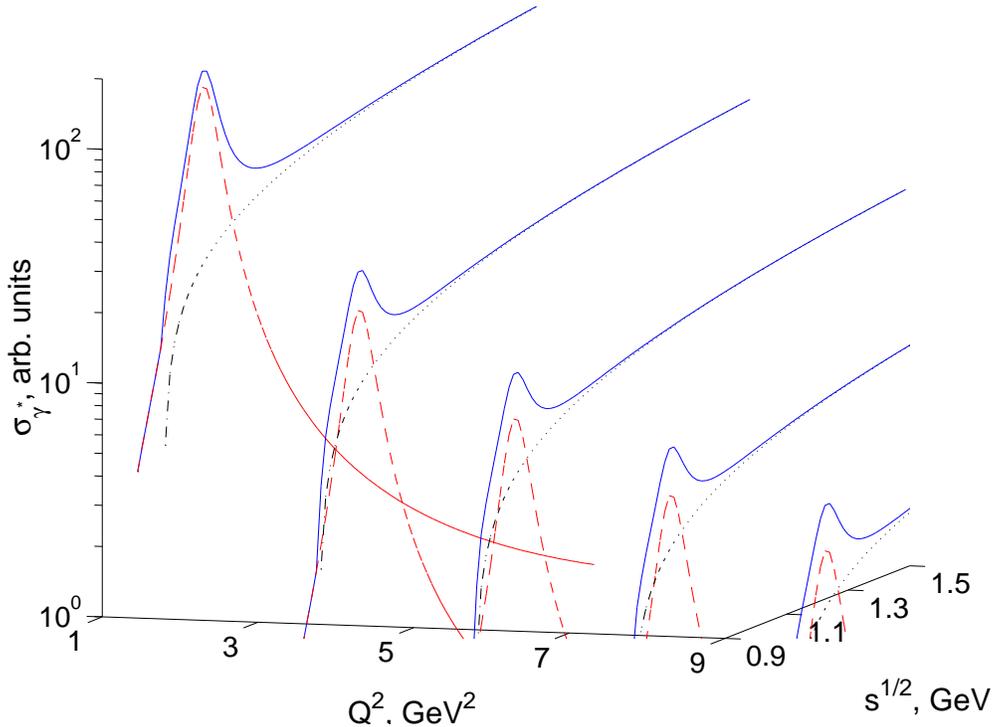}
\caption{ $\gamma^* p$ total cross section as a
function of $\sqrt{s}$ and $Q^2$. For different values of $Q^2$ 
we show the contributions
from the $\Delta$ resonance (dashed line), the background, i.e. 
the contribution from the 
exotic trajectory (dot-dashed line)
and their sum (full line).} \label{fig4}
\end{figure}

In the interval of interest $\sqrt{s}=1.1-1.5\ GeV,\ t=0$ we have
$u=m_N^2-s<0$, so, it is far from resonance region, therefore we
neglect the contribution from $D(u,t)$ for both resonance and
background terms.

The smooth background is also modeled by a single term and exotic
trajectory (\ref{eq32}). As already explained, the direct channel Regge 
pole does not produce here physical resonances. The parameters of the exotic
trajectory are the following: \beq
\alpha_E(s)=-0.25+0.25(\sqrt{1.21}-\sqrt{1.21-s}), \eeq{ex} where
$s_E=1.1^2\ GeV^2$ is an effective exotic threshold. Obviously "pole" 
doesn't mean a resonance in this case.

Figure \ref{fig1}
shows $g$ as a function of $Q^2$ for $\Delta$ and exotic
trajectories.
The resulting cross sections (imaginary part of the amplitude) in
the resonance region is shown in Figs. \ref{fig2} and \ref{fig3}
for two values of $Q^2=1$ and $6\ GeV^2$. It is in qualitative
agreement with the experimental data (compare with Fig. \ref{ress}) \cite{JM,JMP01}. 
Figure \ref{fig4} shows the dual
properties of the cross section in 2 dimensions - one is the squared 
energy $s$ and the other one is virtuality $Q^2$.
Table \ref{tab1} shows the values
of the parameters used in our calculations.  

\section{Conclusions}

The spirit of the present lectures is close to that of the remarkable paper 
\cite{BK}. Moreover, we are trying to construct an explicit model realizing the
continuity of the strong interactions.  We have shown how $Q^2$ dependence can 
be incorporated in DAMA, becoming a function of 3 variables, namely $s,t$ and 
$Q^2$ and thus making complete the circle shown in Fig. \ref{diag}. In particular, 
we answer the question posed at the beginning of the present lectures: "Does the 
small-$x$ structure function know its large-$x$ behavior (and v.v.)?". Our 
answer is affirmative.

The main conclusions from our analysis are that:\\
A) $Q^2$-dependence at low- and high-x (or high- and low-s) are 
interrelated and have the same origin;\\
B) Even a single (low energy) resonance can produce a smooth
scaling-like curve in the structure function (parton-hadron
duality).

To summarize, we have suggested am explicit dual model in which
the $Q^2$-dependence introduced in the low-$x$  domain is extended
to the whole kinematic region, in particular to the region of resonances.
The resulting predictions for the first resonance in the
$\gamma^* p$ system shown in Figs. \ref{fig2},~\ref{fig3} are in
quantitative agreement with data.

\begin{figure*}[htb]
        \insertplotlll{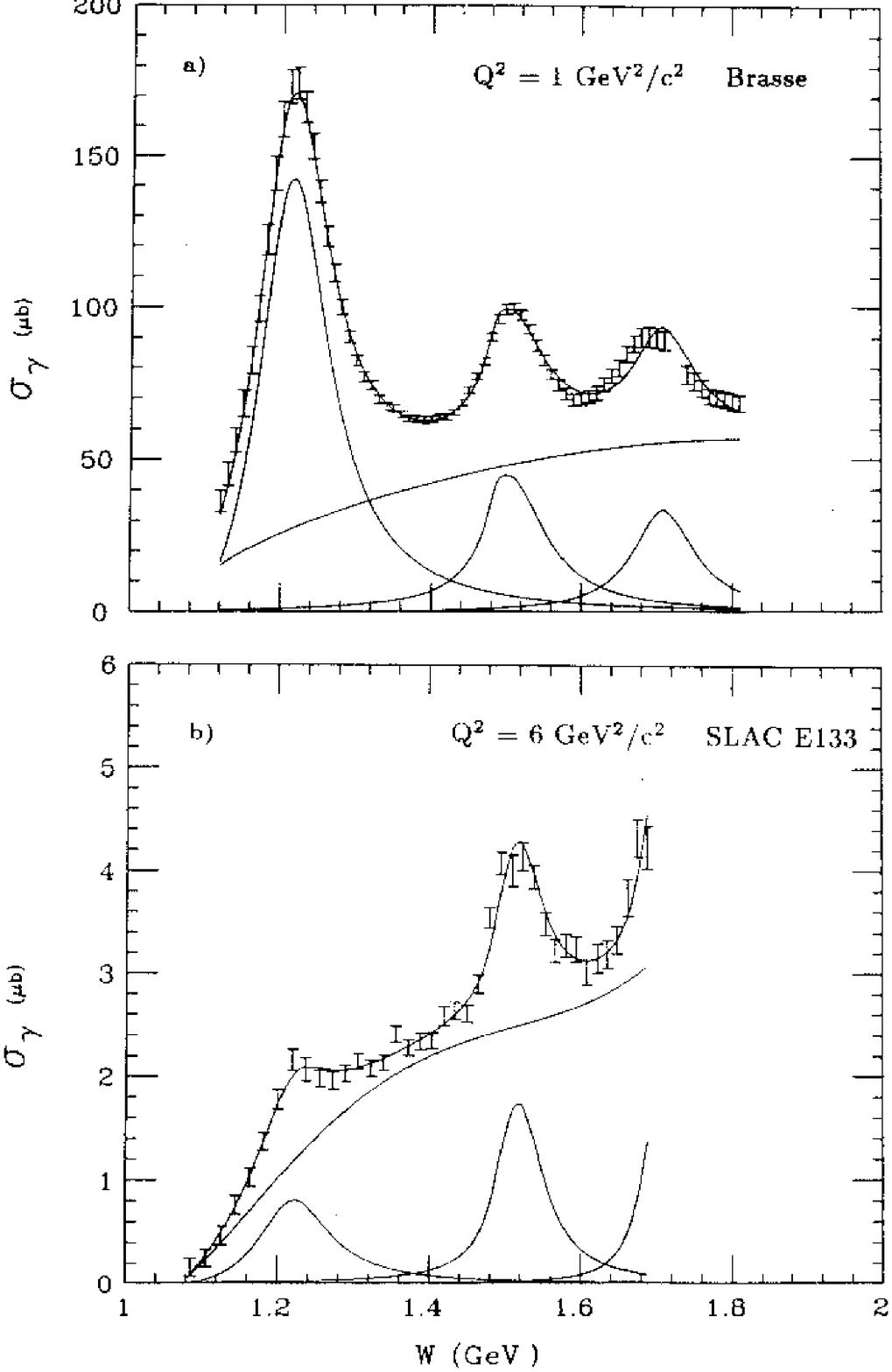}
\vspace*{-0.5cm} 
\caption{Examples of least-squares fits to inclusive $(e,e')$ data. The curves below the data
are the individual resonance and nonresonant contributions.  From \cite{stoler}. } 
\label{ress}
\end{figure*}

\section{Acknowledgments} 
L.J. ans V.M. acknowledge the support 
by INTAS, Grant 00-00366. 
The work of L.J. was supported also by the US Civilian Research and 
Development Foundation (CRDF), Grant UP1-2119.

L.J. thanks the Organizers of the Gomel School-Seminar for the invitation 
and for the warm, pleasant and creative atmosphere during the Seminar and 
Seminar.

\vfill \eject

\end{document}